\documentclass[12pt]{article}
\usepackage{amsmath, amssymb}
\begin{document}
\date{}

\title{The $su(1,1)$ dynamical algebra for the generalized MICZ-Kepler problem from the Schr\"odinger factorization}

\author{M. Salazar-Ram\'{\i}rez$^{a}$, D. Mart\'{\i}nez$^{b}$,\footnote{{\it E-mail address:} dmartinezs77@yahoo.com.mx}\\ V. D. Granados$^{a}$ and R. D. Mota$^{c}$ }\maketitle

\begin{minipage}{0.9\textwidth}
\small $^{a}$Escuela Superior de F\'{\i}sica y Matem\'aticas,
Instituto Polit\'ecnico Nacional,
Ed. 9, Unidad Profesional Adolfo L\'opez Mateos, 07738 M\'exico D. F., M\'exico.\\

\small $^{b}$ Universidad Aut\'onoma de la Ciudad de M\'exico,
Plantel Cuautepec, Av. La Corona 320, Col. Loma la Palma,
Delegaci\'on
Gustavo A. Madero, 07160, M\'exico D. F., M\'exico.\\

\small $^{c}$ Unidad Profesional Interdisciplinaria en Ingenier\'{\i}a y
Tecnolog\'{\i}as Avanzadas, I.P.N., Av. Instituto Polit\'ecnico
Nacional 2580, Col. La Laguna Ticom\'an, Delegaci\'on Gustavo A.
Madero, 07340 M\'exico D. F., M\'exico.\\

\end{minipage}

\begin{abstract}
We apply the Schr\"odinger factorization to construct the
generators of the dynamical algebra $su(1,1)$ for the radial equation of the generalized MICZ-Kepler problem.
\end{abstract}



\section{Introduction}
The factorization methods have played an important role in the
study of quantum systems (Dirac, 1935; Schr\"odinger, 1940, 1940a, 1941; Infeld, 1941; Infeld and Hull, 1951). This is because, if the Schr\"odinger equation is factorizable,
the energy spectrum and the eigenfunctions are obtained algebraically. Infeld and
Hull (Infeld, 1941; Infeld and Hull, 1951) used the ideas of Dirac (1935) and Schr\"odinger (1940, 1940a, 1941) and created a factorization
method (IHFM) which uses a particular solution for the Ricatti equation. This
method allowed to classify the problems according to the characteristics involved in the
potentials and it is closely related to the supersymmetric quantum mechanics (Cooper et al., 1995; Dutt, 1987; Bagchi, 2001; Lahiri, 1990).

The Schr\"odinger factorization (Schr\"odinger, 1940) is a technique which is essentially different to that of the IHFM (Infeld, 1941). In order to clarify these differences we consider a typical central-potential problem like the two-dimensional hydrogen atom or the two-dimensional isotropic harmonic oscillator. The ladder operators obtained by means of the IHFM can only generate a finite number of states with different values of the angular momentum for each degenerate energy eigenvalue (Cordero and Daboul, 2005). Therefore, the symmetry algebras of these problems for bound states are compact. On the other hand, the raising and lowering operators obtained by applying the Schr\"odinger factorization generate an infinite number of states with different values of the energy for each degenerate angular momentum eigenvalue. Thus, dynamical symmetry algebras for these systems are non-compact and have infinite-dimensional representations (Cordero and Daboul, 2005). Historically the Schr\"odinger factorization has been less applied to physical problems than the IHFM.

A systematic method to find both compact and non-compact
algebra generators for a given system has not still developed. These generators have been
intuitively found and forced them to close an algebra, as it is extensively shown in (Bohm et al., 1988; Wybourne, 1974; Englefield, 1972; Mart\'inez-y-Romero et al., 2005; L\'evai, 1994; Wu and Alhassid, 1990; Englefield and Quesne, 1991).
For several one-dimensional potentials it has been shown they exhibit an $su(1,1)$ dynamical algebra (Bagchi, 2001; Mart\'inez-y-Romero et al., 2005; L\'evai, 1994; Wu and Alhassid, 1990; Englefield and Quesne, 1991). It must be emphasize that these authors had to introduce an additional variable to define the generators of this algebra.
On the other hand, in (Mart\'inez and Mota, 2008; Mart\'inez et al., 2009) we have shown that by means of the Schr\"odinger factorization it is possible to construct the generators of the dynamical algebras for some central-potential problems without using any additional variable. 

The MICZ-Kepler problem is the Kepler problem when the nucleus of this hypothetic hydrogen atom also carries a magnetic charge. This problem was independently discovered by McIntosh-Cisneros (1968) and Zwanziger (1968). It has been shown that the MICZ-Kepler problem possesses a Runge-Lenz-type vector as constant of motion and the group $O(4)$ as symmetry group (McIntosh and Cisneros, 1968). These facts reflect a great similarity with the Coulomb problem. Different generalizations for the MICZ-Kepler problem have been studied. For example, Meng (2007) has studied the MICZ-Kepler problem in all dimensions and Mardoyan (2003, 2003a) has solved the Schr\"odinger equation for the generalized MICZ-Kepler Hamiltonian ($\hbar=m=c=1$)         
\begin{equation}
H=\frac{1}{2}\left(-i\nabla-s\bold{A}\right)^2+\frac{s^2}{2r^2}-\frac{1}{r}+\frac{c_1}{r(r+z)}+\frac{c_2}{r(r-z)},
\label{EQ1}
\end{equation}        
where $\bold A$ is the magnetic vector potential of a Dirac monopole,
\begin{equation}
{\bold A}=\frac{1}{r(r-z)}(y,-x,0),
\end{equation}
such that $\nabla\times {\bold A}=\frac{{\bold r}}{r}$, $r$ is the distance from the electron to the hydrogen nucleus, $s$ is the magnetic charge of the Dirac monopole which takes the values $s=0$, $\pm\frac{1}{2},\pm 1,\pm \frac{3}{2}$... and,  $c_1$  and $c_2$ are non-negative constants. Moreover, there are generalizations of the MICZ-Kepler systems on the three-dimensional sphere (Gristev et al., 2000) and on the hyperboloid (Nersessian and Pogosyan, 2001). Recently,  Giri studied the radial Schr\"odinger equation corresponding to the generalized MICZ-Kepler problem (equation (\ref{EQ1})) from supersymmetric quantum mechanics (Giri, 2008). In this way he obtained the superpotential from which the SUSY factorization operators are defined. It is immediate to show that the action of these operators on the radial eigenstates $R_{nj}(r)$ of the Hamiltonian (\ref{EQ1}) is to change the secondary quantum number $j$ leaving the principal quantum number $n$ fixed.

In this work we apply the Schr\"odinger factorization (Schr\"odinger, 1940; Mart\'inez and Mota, 2008) to the MICZ-Kepler problem and obtain ladder operators which change the energy quantum number leaving fixed the total angular momentum quantum number. Because of what we emphasized above, these operators must be related to the algebras of the non-compact symmetry groups. In fact, we show how the Schr\"odinger operators allow us to construct the generators of the $su(1,1)$ dynamical algebra for the bound states of the radial equation for the generalized MICZ-Kepler Hamiltonian (\ref{EQ1}). The paper is organized as follows. In section 2, we summarize the important points of the Mardoyan papers (2003, 2003a) which are relevant to this work. In section 3, the Schr\"odinger factorization is performed and the $su(1,1)$ dynamical algebra is constructed. Finally, in section 4, we give the concluding remarks.

\section{The generalized MICZ-Kepler problem in spherical basis}

The Schr\"odinger equation $H\Psi=E\Psi$ for the Hamiltonian (\ref{EQ1}) with $\Psi\equiv R(r)Z(\theta,\phi)$ in spherical coordinates $(r,\theta,\phi)$, can be reduced to the uncoupled differential equations (Mardoyan, 2003, 2003a; Giri, 2008)   
\begin{align}
&\frac{1}{\sin\theta}\frac{\partial}{\partial \theta}\left(\sin\theta\frac{\partial Z}{\partial \theta}\right)+\frac{1}{4\cos^2\frac{\theta}{2}}\left(\frac{\partial^2}{\partial\phi^2}-4c_1\right)Z+\frac{1}{4\sin^2\frac{\theta}{2}}\left[\left(\frac{\partial}{\partial\phi}+2is\right)^2-4c_2\right]Z=-{\cal A}Z,\label{ANG}\\
&\frac{1}{r^2}\frac{d}{dr}\left(r^2\frac{dR}{dr}\right)-\frac{{\cal A}}{r^2}R+2\left(E+\frac{1}{r}\right)R=0, \label{RAD}
\end{align}
where the quantized separation constant is given by
\begin{equation}
{\cal A}=\left(j+\frac{\delta_1+\delta_2}{2}\right)\left(j+\frac{\delta_1+\delta_2}{2}+1\right). 
\label{QCON}
\end{equation}  
The unnormalized square-integrable solutions for the equation (\ref{ANG}) are (Mardoyan, 2003, 2003a) 
\begin{equation}
Z^{(s)}_{jm}(\theta,\phi;\delta_1\delta_2)=\left(\cos\frac{\theta}{2}\right)^{m_1} \left(\sin\frac{\theta}{2}\right)^{m_2}P^{(m_2,m_1)}_{j-m_+}(\cos\theta)e^{i(m-s)\phi},\label{ASOL}
\end{equation}
where $m_1=|m-s|+\delta_1=\sqrt{(m-s)^2+4c_1}$,$m_2=|m+s|+\delta_2=\sqrt{(m+s)^2+4c_2}$, $m_+=(|m+s|+|m-s|)/2$ and $P^{(a,b)}_n$ are the Jacobi polynomials. The $z$-component of the total angular momentum $m$ and the total angular momentum $j$ take the quantized eigenvalues
\begin{align}
&m=-j,-j+1,...,j-1,j\\
&j=\frac{|m+s|+|m-s|}{2}, \frac{|m+s|+|m-s|}{2}+1,....\label{j}
\end{align}  
Notice that the Dirac quantization condition $s$ determines $j$ and $m$. The equations above imply that $j$ and $m$ take integer or half-integer values depending on whether $s$ takes integer or half-integer values.        

By substituting the quantized separation constant (\ref{QCON}) into equation (\ref{RAD}) we obtained
\begin{equation}
\frac{1}{r^2}\frac{d}{dr}\left(r^2\frac{dR}{dr}\right)-\frac{1}{r^2}\left(j+\frac{\delta_1+\delta_2}{2}\right)\left(j+\frac{\delta_1+\delta_2}{2}+1\right)R+2\left(E+\frac{1}{r}\right)R=0. \label{RAD2}
\end{equation}
This equation under the change  $j+(\delta_1+\delta_2)/2 \rightarrow 
J$ results to be equal to the radial equation for the hydrogen atom (Davidov, 1976). Thus, the solutions for discrete spectrum of equation (\ref{RAD2}) apart from normalization are (Mardoyan, 2003, 2003a)
\begin{equation}
R^{(s)}_{nj}=(2\epsilon r)^{j+\frac{\delta_1+\delta_2}{2}}e^{-\epsilon r}F(-n+j+1,2j+\delta_1+\delta_2+2;2\epsilon r),
\end{equation} 
where $n=|s|+1|, |s|+2,...$, $F(a,b;z)$ is the confluent hypergeometric function, and   
the parameter $\epsilon$ is given by
\begin{equation}
\epsilon=\sqrt{-2E_n}=\frac{1}{n+\frac{\delta_1+\delta_2}{2}},
\end{equation}
from which the exact energy spectrum 
\begin{equation}
E_n\equiv E_n^{(s)}=-\frac{1}{2\left(n+\frac{\delta_1+\delta_2}{2}\right)^2}
\label{SPEC}
\end{equation}
is obtained.

Mardoyan (2003, 2003a) also showed that the Schr\"odinger equation for the MIC-Kepler system is separated in parabolic coordinates.

\section{The Schr\"odinger ladder operators and the $su(1,1)$ dynamical algebra}

Performing the change $R_{nj}(r)=\chi_{nj}(r)/r$ on the equation (\ref{RAD2}), we obtain 
\begin{equation}
\left(-r^2\frac{d^2}{dr^2}-2r-2E_nr^2\right)\chi_{nj}(r)=-\left(j+\frac{\delta_1+\delta_2}{2}\right)\left(j+\frac{\delta_1+\delta_2}{2}+1\right)\chi_{nj}(r).
\label{MAS1}
\end{equation}
We set 
\begin{equation}
K^2_n=-\frac{1}{2E_n}\label{k}
\end{equation} 
and $r=K_nx$ to write this equation in the form
\begin{equation}
{\cal L}_n\chi_{nj}\equiv \left(-x^2\frac{d^2}{dx^2}-2K_nx+x^2\right)\chi_{nj}=-J(J+1)
\chi_{nj},
\label{MAS2}
\end{equation}
where $J\equiv j+\frac{\delta_1+\delta_2}{2}$.

In order to factorize the operator ${\mathcal L}_n$ we apply the Schr\"odinger factorization (Schr\"odinger, 1940; Mart\'inez and Mota, 2008). Thus, we propose a pair of first-order differential operators such that
\begin{equation}
\left(x\frac{d}{dx}+ax+b\right)\left(-x\frac{d}{dx}+cx+f\right)\chi_{nj}=g\chi_{nj},
\label{sch}
\end{equation}
where  $a$, $b$, $c$, $f$ y $g$ are constants to be determined. Expanding this expression and comparing it with equation (\ref{MAS2}) we obtain
\begin{equation}
a=c=\pm1,\hspace{2ex} b=f-1=\mp K_n-1,\hspace{2ex} g=K_n(K_n\pm1)-J(J+1).
\end{equation}
Using these results equation (\ref{MAS2}) is equivalent to 
\begin{eqnarray}
(T_-^{n}-1)T_+^n \chi_{nj}=\left[K_n(K_n+1)-J(J+1)\right]\chi_{nj},\label{tn1}\\
(T_+^n+1)T_-^n \chi_{nj}=\left[K_n(K_n-1)-J(J+1)\right]\chi_{nj},\label{tn2}
\end{eqnarray}
where we have defined the operators 
\begin{equation}
T^n_\pm=\mp x\frac{d}{dx}+x-K_n.
\label{Tn}
\end{equation}

If we define the operator
\begin{equation}
T_3=\frac{1}{2}\left(-x\frac{d^2}{dx^2}+x+\frac{J(J+1)}{x}\right)
\label{D3}
\end{equation}
and considering equation (\ref{MAS2}) we have
\begin{equation}\label{TtresMIC}
T_3\chi_{nj}=K_n\chi_{nj}.
\end{equation}

Therefore, this operator allows to define the new operators 
\begin{equation}
T_{\pm}=\mp
x\frac{d}{dx}+x-T_3. \label{T+-}
\end{equation}
A direct calculation shows that the set of operators $T_{\pm}$ and $T_{3}$ satisfy the commutation relations
\begin{eqnarray}\label{opmasmen}
[T_\pm,T_3]& =\mp T_\pm,\\
\lbrack T_+,T_-\rbrack& =-2T_3.
\end{eqnarray}
Thus, the Schr\"odinger factorization allowed us to construct the
generators of the $su(1,1)$ Lie algebra for the generalized
MICZ-Kepler problem. 

The quadratic Casimir operator
\begin{equation}
T^2\equiv -T_{\pm}T_\mp+{T_3}^2{\mp}T_3,
\end{equation}
satisfies the eigenvalue equation
\begin{equation}
T^2\chi_{nj}=J\left(J+1\right)\chi_{nj} .\label{CasEn}
\end{equation}

In order to obtain the energy spectrum, we consider the theory of unitary irreducible representations of the $su(1, 1)$
Lie algebra, which has been studied in several works (Adams, 1988) and it is based on the eigenvalue equations
\begin{align}
C^2\Psi_{\nu\mu}&=\mu(\mu+1)\Psi_{\nu\mu},\label{c2}\\
C_3\Psi_{\nu\mu}&=\nu\Psi_{\nu\mu}\label{c3}.
\end{align}
where $C^2$ is the quadratic Casimir operator, $\nu=\mu+n'+1$,
$n'=0,1,2,...$ and $\mu>-1$. 

From equations (\ref{CasEn}) and (\ref{c2}) we find $\mu=J$. Thus
\begin{equation}
\nu=J+n'+1=j+\frac{\delta_1+\delta_2}{2}+n'+1. \label{EnerK}
\end{equation}
If we define $n\equiv j+n'+1$ and by using equations (\ref{k}), (\ref{TtresMIC}), (\ref{c3}) and (\ref{EnerK}) we find that the energy spectrum for the MICZ-Kepler problem is
\begin{equation}
E_n=-\frac{1}{2\left(n+\frac{\delta_1+\delta_2}{2}\right)^2}.
\end{equation}
From equation (\ref{j}), $j$ takes integer or half-integer values, hence $n$ takes integer or half-integer values, which is in accordance with the result obtained by using the confluent hypergeometric function, equation (\ref{SPEC}) .

Equations (\ref{TtresMIC}) and (\ref{opmasmen}) allow to obtain
\begin{equation}\label{RelKn}
T_3T_\pm\chi_{nj}=\left(K_n\pm1\right)T_\pm\chi_{nj}.
\end{equation}
From (\ref{TtresMIC}), (\ref{c3}) and (\ref{EnerK}), and the definition of $n$, we obtain that $K_n$ satisfies the relation
\begin{equation}
K_{n\pm1}=K_n\pm1.\label{K}
\end{equation}
Thus, using relation (\ref{K}), we show
\begin{equation}
T_\pm\chi_{nj}\propto \chi_{n\pm 1\,j}.\label{asdes3}
\end{equation}
From this expression and equations (\ref{TtresMIC}) and (\ref{T+-}), we find that the Schr\"odinger operators, $T_\pm^n$, acting on the eigenstates $\chi_{nj}$ satisfy
\begin{equation}
T_\pm^n\chi_{nj}\propto \chi_{n\pm 1\,j}.
\end{equation}
In this way we showed the essence of the Schr\"odinger factorization which is to find operators that relate eigenstates belonging to the same secondary quantum number $j$ but different principal quantum number $n$. It must be pointed out that these results are achieved without the knowledge of the explicit form of the eigenfunctions.

\section{Concluding Remarks}

We have applied the Schr\"odinger factorization to the radial equation of the
generalized MICZ-Kepler problem and a pair of first-order differential operators
were obtained.  From the radial Hamiltonian we introduced, in a
natural way, a third operator which closes the $su(1,1)$  dynamical algebra for this problem. Moreover, we obtained from a purely algebraic way the energy spectrum and showed that the action of the Schr\"odinger operators on the radial eigenstates is to change only the principal quantum number $n$ leaving fixed the total angular momentum quantum number $j$. Notice that for the construction of these generators we did not introduced any additional variable, which differs form the procedure follow in (Mart\'inez-y-Romero et al., 2005; L\'evai, 1994; Wu and Alhassid, 1990; Englefield and Quesne, 1991). To our knowledge the results we have shown in this paper have not
been found before.

If we restrict our results to the standard MICZ-Kepler problem, then
the $O(4)$ symmetry of this problem accounts for the degenerate states for a fixed energy, whereas the $su(1,1)$ dynamical algebra accounts for the degenerate states for fixed total angular momentum. As it is well known the  hydrogen atom possesses the $su(1,1)$ dynamical symmetry (Wybourne, 1974). Therefore,  the existence of the $su(1,1)$ dynamical algebra  
and a Runge-Lenz type vector as constant of motion for the MICZ-Kepler problem  makes this system completely similar to the non-relativistic hydrogen atom, at least at algebraic level.

As another application of our method we have studied the relativistic hydrogen atom for which we have constructed the three generators for the $su(1,1)$ (Salazar-Ram\'irez et. al., 2010).

\section*{Acknowledgments}

This work was partially supported by SNI-M\'exico, CONACYT grant
number J1-60621-I, COFAA-IPN, EDI-IPN, SIP-IPN projects numbers
20091042 and 20090590, and ADI-UACM project numbers 7D92023009 and 7DA2023001.

\section*{References}

\begin{list}{}{}
\item Adams, B. G., {\it et. al.}, Adv. Quant. Chem. Vol. 19 (1987)1 reprinted in  Bohm, A., {\it et. al.}, Dynamical Groups and Spectrum Generating Algebras, vols. 1 and 2, World Scientific, Singapore, 1988
\item Bagchi, B.K., Supersymmetry in Quantum and Classical Mechanics, Chapman \& Hall/CRC, USA, 2001
\item Bohm, A., Ne'eman, Y., Barut, A.O., Dynamical Groups and Spectrum Generating Algebras, Vols. 1 and 2, World Scientific, Singapore, 1988
\item Cooper, F., Khare, A., Sukhatme, U., Phys. Rep. {\bf251}, 267-385(1995)
\item Cordero, P., Daboul, J., J. Math.  Phys. {\bf46}, 053507(2005)
\item Davidov, A.S., Quantum Mechanics, Pergamon, Oxford, 1976
\item Dirac, P. A. M., The Principles of Quantum Mechanics, Clarendon Press, Oxford, 1935
\item Dutt, R., Khare, A., Sukatme, U.P., Am. J. Phys. {\bf56}, 163(1987)
\item Englefield, M.J., Group Theory and the Coulomb Problem, Wiley Interscience, USA, 1972
\item Englefield, M. J.,  Quesne, C., J. Phys. A: Math. Gen. {\bf24}, 3557(1991)
\item Giri, P.R., Mod. Phys. Lett. A {\bf23}, 895(2008)
\item Gritsev, V. V., Kurochkin, Yu. A., Otchik, V. S., J. Phys. A: Math. and Gen. {\bf33}, 4903(2000)
\item Infeld, L., Phys. Rev. {\bf59}, 737(1941)
\item Infeld, L., Hull, T.E., Rev. Mod. Phys. {\bf23}, 21(1951)
\item Lahiri, A., Roy, P.K., Bagchi, B., Int. J. Mod. Phys. A {\bf5}, 1383-1456(1990)
\item L\'evai, G., J. Phys. A: Math. Gen. {\bf27}, 3809(1994)
\item Mardoyan, L., J. Math.  Phys. {\bf44}, 4981(2003)
\item Mardoyan, L., quant-p/0310143v1(2003a)
\item Mart\'{\i}nez, D., Mota, R. D., Ann. Phys. {\bf323}, 1024(2008)
\item Mart\'{\i}nez, D., Flores-Urbina, J. C., Mota, R. D., Granados, V. D., to be published in J. Phys. A: Math. and Theor.,(2009)
\item Mart\'{\i}nez-y-Romero, R.P., N\'u{\~n}ez-Y\'epez, H.N., Salas-Brito, A.L., J. Phys. A: Math. Gen. {\bf38}, 8579(2005)
\item McIntosh, H. V., Cisneros, A., J. Math.  Phys. {\bf11}, 896(1968)
\item Meng, G. W., J. Math.  Phys. {\bf48}, 320(2007), hep-th/0607059v3
\item Nersessian, A., Pogosyan, G., Phys. Rev. A {\bf63}, 020103(R)(2001)
\item Salazar-Ram\'{\i}rez, M., Mart\'{\i}nez, D., Granados, V.D., Mota, R. D., submitted to Phys. Lett. A.
\item Schr\"odinger, E., Proc. R. Ir. Acad. A {\bf46}, 9(1940)
\item Schr\"odinger, E., Proc. R. Ir. Acad. A {\bf46}, 183(1940a)
\item Schr\"odinger, E., Proc. R. Ir. Acad. A {\bf47}, 53(1941)
\item Wu, J., Alhassid, Y., J. Math. Phys. {\bf31}, 557(1990)
\item Wybourne, B.G., Classical Groups for Physicists, Wiley Interscience, New York, 1974
\item Zwanziger, D., Phys. Rev. {\bf176}, 1480(1968)
\end{list}

\end{document}